\documentclass[12pt]{article}
\usepackage[dvips]{color}
\usepackage{epsfig}
\usepackage{amsmath}
\usepackage{cite}
\usepackage{graphicx}
\usepackage{color}
\usepackage{hyperref}
\usepackage{subfigure}

\begin{document}
\begin{center}
\Large{\bf Weak Gravity Conjecture, Black Branes and Violations of Universal Thermodynamic Relation}\\
\small \vspace{1cm} {\bf J. Sadeghi$^{a}$\footnote {Email:~~~pouriya@ipm.ir}}, \quad
{\bf B. Pourhassan$^{b}$\footnote {Email:~~~b.pourhassan@du.ac.ir}}, \quad
{\bf S. Noori Gashti$^{a}$\footnote {Email:~~~saeed.noorigashti@stu.umz.ac.ir}}, \quad
{\bf S. Upadhyay$^{d,b}$\footnote {Email:~~~sudhakerupadhyay@gmail.com}}, \quad
\\
\vspace{0.5cm}$^{a}${Department of Physics, Faculty of Basic Sciences,\\
University of Mazandaran P. O. Box 47416-95447, Babolsar, Iran}\\
\vspace{0.5cm}$^{b}${School of Physics, Damghan University, Damghan, 3671641167, Iran}\\
\vspace{0.5cm}$^{d}${Department of Physics, K. L. S. College, Nawada, Bihar 805110, India}
\\
\vspace{0.5cm}$^{d}${Department of Physics, Magadh University, Bodh Gaya,
 Bihar  824234, India}\\
\vspace{0.5cm}$^{d}${Inter-University Centre for Astronomy and Astrophysics (IUCAA), Pune, Maharashtra 411007, India}
\\
\small \vspace{1cm}
\end{center}
\begin{abstract}
The universal thermodynamic relations between corrections to entropy and extremality  for various black holes solutions have been studied. In this regard, we hereby consider a number of different black brane solutions  in different structures  for perturbative corrections to general relativity. These are, namely,
  black brane solution in Rastall AdS massive gravity, Einstein-Yang-Mills AdS black brane solution in massive gravity and general anisotropic black brane in Horava-Lifshitz gravity. We calculate  both the entropy and extremality bound by introducing
   a small constant correction to the action. Remarkably, we find  that black brane
       violates the universal thermodynamic relations. In other words, a universal relation between corrections to entropy and extremality  are not valid in the black brane structure.
\end{abstract}
\newpage
\section{Introduction and Motivation}
 General Relativity is considered as a low-energy effective field theory (EFT) of the gravity. The impact of quantum gravity   on low-energy EFTs is important to study  for a regime where quantum gravity  become significant.   Despite being treated often as classical objects, black hole solutions have provided remarkable insights into quantum gravity. The swampland program was introduced  to estimate   constraints that quantum gravity puts on EFTs \cite{a, c,d}.
 In this connection, weak gravity conjecture (WGC) is one of constraints which
 describes that any EFT with a U(1) gauge field coupled to gravity must have charge state greater than mass   \cite{e}.

Black hole singularities  hidden by horizons  must have  charge
state to be less than mass and imposes ``extremality bounds" for many second-derivative theories. Moreover, higher-derivative corrections can change such bounds, and the  charge state  of the black hole can be more or less than its mass.
In Refs. \cite{f,g},  it is found that  corrections to the extremality bound in a canonical ensemble are related to corrections to the entropy in a microcanonical ensemble.  Goon and Penco \cite{1}  reformulates the WGC and found that this entropy/extremality relationship (universal relation) follows quite general from thermodynamics.

 The universal thermodynamic relation has since been explored extensively and various implications of these relations are studied under different conditions  \cite{2,3,4,5,6,7,8,9}. For instance, the universal relation has   been studied for a Kerr-Newman-AdS black hole surrounded by quintessence and the cloud of string
  \cite{10}  and AdS black holes surrounded by perfect fluid dark matter \cite{11}. Generally, this universal thermodynamic relation has always been valid for black holes with different characteristics and conditions. Such studies motivated us to look at these universal relationships from a new perspective, which led to exciting developments. Here, we consider different structures of the black brane, in particular, solution in Rastall AdS massive gravity \cite{12}, Einstein-Yang-Mills AdS black brane solution in massive gravity \cite{13}, and general anisotropic black brane in Horava-Lifshitz gravity \cite{14}. Then, we try to examine a different form of these universal relations. To give a general description of a black brane in general relativity, we emphasize that the solution of equations that generalizes a black hole solution has also been extended to additional spatial dimensions. Different descriptions can be given for it, for e.g., in string theory, the black brane describes a group of D1 branes surrounded by a horizon. By recognizing the concept of horizons and points as zero branes, the generalization of a black hole is
  made to a black p-brane \cite{15,16,17,18,19,20}.

The purpose of this paper is to investigate the new implications of black brane with respect to the universal thermodynamic relations. Therefore, considering the three different structures of the black brane, regarding the universal thermodynamic relations, we examine the results and first obtain the modified thermodynamic relations by considering a small constant correction added to the black brane action. Then, by analytical calculations, we investigate these universal relations
and incidently find  that  these universal relations are not valid in general for different black branes with various properties and structures. In fact, under certain
approximation, such relations may hold for black branes as well.
 However, these universal relations hold strongly for black hole solutions.

All the above information motivated us to organize the paper as following. In sections \ref{sec2}, \ref{sec3}, and \ref{sec4}, we obtain the universal relations for black brane solution in Rastall AdS massive gravity, Einstein-Yang-Mills AdS black brane solution in massive gravity, and general anisotropic black brane in Horava-Lifshitz gravity, respectively, in accordance with the concepts mentioned. It becomes clear that these universal thermodynamic relations are not valid for the general black brane solutions. So a black brane system may violate such universal relation except for the special cases. Finally, in section \ref{sec5}, we conclude the results of our work.

\section{Perturbative  Black Brane Solution in Rastall AdS Massive Gravity}\label{sec2}
 According to general relativity, the covariant divergence of the energy-momentum vanishes. However, such conservation laws fulfils certain limitation that theory is testable only   in flat space. The curved space-time   generalization could not be
 easily done. In order to overcome this situation, non-Lagrangian Rastall theory is proposed which modifies  such conservation laws by the form $T^{\mu}_{\mu,\nu}=\lambda R_{,\nu}$, where $\lambda$ is the Rastall parameter \cite{21}. For vanishing $\lambda$, this reduces to the original   Einstein  field equations. The Lagrangian of this theory is   written by Smalley  \cite{22} who extracted the field equations with the help of variational principle. Fierz-Pauli introduced massive gravity in flat space-time \cite{24}. The massive gravity    is constructed  to curved space-time which are ghost free in the decoupling limit to all orders in Ref. \cite{23}. The action for  Rastall theory in massive gravity  is given by
\begin{equation}\label{1}
I=\int d^{4}x\sqrt{-g}\left[\frac{1}{2k^{'}} e^{ 2\sqrt{-g}\lambda^{'}k^{'} }R -2\Lambda +m^{2} \sum_{i=1}^{4}c_{i}U_{i}(g,f)\right],
\end{equation}
where   $\lambda^{'}$ and $k^{'}$ are constant parameters \cite{12,22}.
Here,  $c_i$  are constants and $U_i$ are symmetric polynomials of the eigenvalues. The Rastall field equation in massive gravity  is given by
\begin{equation}\label{2}
R_{\mu\nu}-\frac{1}{2}(R-2\Lambda)g_{\mu\nu}+k\lambda g_{\mu\nu}R-m^{2}\chi{\mu\nu}=kT_{\mu\nu},
\end{equation}
where
  $R$ is  Ricci scalar,  $k=k'e^{-2\sqrt{-g}\lambda^{'}k^{'} }$ and $\lambda=\lambda'e^{-2\sqrt{-g}\lambda^{'}k^{'} }$  are  covariantly constant parameters  \cite{12}.  The Einstein and  Rastall  tensors are given, respectively, by
\begin{equation}\label{3}
G_{\mu\nu}=R_{\mu\nu}-\frac{1}{2}Rg_{\mu\nu},
\end{equation}
 \begin{equation}\label{4}
H_{\mu\nu}=G_{\mu\nu}+k\lambda g_{\mu\nu}R.
\end{equation}
The black brane solution for the above theory is given by
\begin{equation}\label{5}
ds^{2}=-f(r)dt^{2}+\frac{dr^{2}}{f(r)}+\frac{r^{2}}{l^{2}}(dx^{2}+dy^{2}),
\end{equation}
where metric function has the following form:
\begin{equation}
 f(r)=1-\frac{b}{r}-\frac{\Lambda}{3}r^{2}+m^{2}l^{2}\left(\frac{c_{0}c_{1}}{2}r+c_{0}^{2}c_{2}\right).\label{6}
\end{equation}
The vanishing metric function ($f(r)|_{r=r_+} =0$)  gives the event horizon and can be used to found the parameter $b$. The cosmological constant and Rastall parameter
are related as   $\Lambda=\frac{1}{3}\frac{\rho_{0}}{4\lambda-1}$ \cite{12,26}.

The Hawking temperature and the entropy can be calculated, respectively, using Hawking-Bekenstein relation as follows
\begin{equation}\label{7}
T=\frac{1}{4\pi}f'(r_{+})=\frac{1}{4\pi}\left(\frac{1}{r_{+}}-\Lambda r_{+}+\frac{m^{2}l^{2}c_{0}^{2}c_{2}}{r_{+}}+m^{2}l^{2}c_{0}c_{1}\right),
\end{equation}
and
\begin{equation}\label{8}
S=\frac{A}{4G}=\frac{r_{+}^{2}V}{4l^{2}G}.
\end{equation}
Here, $V$ is volume of the constant hyper-surface with $\frac{1}{4G}=4\pi$.

Since General Relativity is a low-energy  effective theory and low-energy signatures of its UV completions are described by higher-derivative corrections. For macroscopically large objects such as  black holes, these additional operators perturbatively correct black hole states and their derived properties such as the Hawking temperature, entropy, and extremality bounds.
Now, in order to have a  perturbatively changing  theory in a manner controlled by a parameter   $\epsilon$, we write action as follows
\begin{eqnarray}\label{9}
I_\epsilon &=&\int d^{4}x\sqrt{-g}\left[\frac{1}{2k^{'}}  e^{ 2\sqrt{-g}\lambda^{'}k^{'} }R- 2(1+\epsilon)\Lambda
+ m^{2} \sum_{i=1}^{4}c_{i}U_{i}(g,f)\right].
\end{eqnarray}
This leads to   modification in the Einstein's equations, and  therefore, the black brane solution of the theory, which has always been used to justify or explain certain phenomena or even acquire new concepts. Corresponding to  above corrected action, the modified thermodynamic quantities such as mass and temperature are obtained, respectively, as
\begin{eqnarray}
M&=&\frac{-2lS\Lambda+l^{3}m^{2}S(1+\epsilon)c_{0}c_{1}+2\sqrt{\pi S V}(1+S+l^{2}m^{2}c_{2}^{2}c_{2})}{4l\pi(1+\epsilon)},\label{10}\\
T&=&\frac{\sqrt{\pi V}(1+\epsilon)-2l\sqrt{S}\Lambda+l^{2}m^{2}c_{0}(l\sqrt{S}(1+\epsilon)c_{1}+\sqrt{\pi V}c_{0}c_{2})}{4l\pi\sqrt{S}(1+\epsilon)}.\label{11}
\end{eqnarray}
Corresponding to Eq. (\ref{10}), the correction parameter can have the following form:
\begin{equation}\label{12}
\epsilon=\frac{4lM\pi-2\sqrt{\pi SV}+2lS\Lambda-l^{3}m^{2}Sc_{0}c_{1}-2l^{2}m^{2}\sqrt{\pi SV}c_{0}^{2}c_{2}}{-4lM\pi+2\sqrt{\pi SV}+l^{3}m^{2}c_{0}c_{1}}.
\end{equation}
The  mass can naturally be written as $M=M_0(S)+\epsilon\Delta M(S)$.
The derivatives of (\ref{10}) and (\ref{11}) determine
\begin{eqnarray}
-T\left(\frac{\partial S}{\partial \epsilon}\right)_M&=&\frac{S\Lambda-lm^{2}\sqrt{\pi S V}c_{0}^{2}c_{2}}{2\pi(1+\epsilon)^{2}}\nonumber\\
&=&\frac{\sqrt{V}(1+\epsilon+l^{2}m^{2}c_{0}^{2}c_{2})\left(-lm^{2}\sqrt{V}c_{0}^{2}c_{2}+ \frac{\Lambda\sqrt{V}(1+\epsilon+l^{2}m^{2}c_{0}^{2}c_{2}) }{(-2l\Lambda+l^{3}m^{2}(1+\epsilon)c_{0}c_{1}) }\right)}{2(-2l\Lambda+l^{3}m^{2}(1+\epsilon)c_{0}c_{1})(1+\epsilon)^{2}}.\label{14}
\end{eqnarray}
In order to study universal relation, we need to evaluate (\ref{14})  at the extremal point $M=M_{ext}(\epsilon$) corresponding to $T=0$ in the corrected theory.

The extremal mass is calculated by plugging the value of entropy in Eq. (\ref{10})
as follows
\begin{eqnarray}
M_{ext}&=& \frac{\sqrt{V}(1+\epsilon+l^{2}m^{2}c_{0}^{2}c_{2})}{4l^{2}(1+\epsilon)(-2\Lambda+l^{2}m^{2}(1+\epsilon)c_{0}c_{1})}\left[
\sqrt{V}(1+\epsilon)  -4 \Lambda\sqrt{V}\frac{(1+\epsilon+l^{2}m^{2}c_{0}^{2}c_{2}) }{  2 \Lambda -l^{2}m^{2}(1+\epsilon)c_{0}c_{1}  } \right.\nonumber\\
&+&\left. l^2m^{2}c_{0}\sqrt{V}\left(c_{0}c_{2}-2 (1+\epsilon)c_{1} \frac{(1+\epsilon +l^{2}m^{2}c_{0}^{2}c_{2}) }{  2 \Lambda -l^{2}m^{2}(1+\epsilon)c_{0}c_{1}   }\right) \right]. \label{15}
\end{eqnarray}
  By taking derivative of  (\ref{15}) with respect to $\epsilon$ and simplifying the
  result, we observe that result does not coincide  with (\ref{14}), i.e.
\begin{equation}\label{16}
\frac{\partial M_{ext}}{\partial\epsilon}\neq -T\left(\frac{\partial S}{\partial \epsilon}\right)_M.
\end{equation}
 From the equation (\ref{16}), we see that the Goon-Penco universal extremality relation proved for black holes is not valid for this black brane.
However, the exciting point here is that  for coefficients $c_{0}=c_{1}=c_{2}=0$  the  universal relation holds for these models as well. In this case, we have
\begin{equation*}
\frac{\partial M_{ext}}{\partial\epsilon}=-T\frac{\partial S}{\partial\epsilon}=\frac{V}{8l^{2}\Lambda}.
\end{equation*}
Entropy $S$, angular momentum $J$, and charge $Q$ can be considered as the parameters for the mass $M(S, J, Q)$. Therefore, temperature, angular velocity, and electric potential  can be expressed as follows
\begin{equation}\label{17}
T=\frac{\partial M}{\partial S}, \hspace{1cm}\Omega=\frac{\partial M}{\partial J}, \hspace{1cm} \Phi=\frac{\partial M}{\partial Q},
\end{equation}
and  in such case another universal relation that holds for black holes can be examined with respect to equations (\ref{10}), (\ref{11}), and (\ref{12}) \cite{2,3,4,5,6,7,8,9}.

\section{Perturbative Einstein-Yang-Mills AdS Black Brane in Massive Gravity}\label{sec3}
Here,  we consider a $5$-dimensional Einstein-Yang-Mills AdS black brane solution in massive gravity as another model of the black brane to identify the correctness  of universal thermodynamic relations. So, let us begin with the action for Einstein-massive gravity in the presence of Yang-Mills source and cosmological constant   as follows
\begin{equation}\label{18}
I=\int d^{5}x\sqrt{-g}(R-2\Lambda-\gamma_{ab}F^{a}_{\mu\nu}F^{b\mu\nu}+m^{2}\Sigma_{i=1}^{4}c_{i}U_{i}(g, f)),
\end{equation}
where $R$, $\Lambda$ and $F^{a}_{\mu\nu}$ are the Ricci scalar, cosmological constant and  $SO(5, 1)$ Yang-Mills gauge field tensor, respectively \cite{13,27}.
Here, $F^{a}_{\mu\nu}$  is defined as
\begin{equation}\label{19}
F^{a}_{\mu\nu}=\partial_{\mu}A_{\nu}^{a}-\partial_{\nu}A_{\mu}^{a}+\frac{1}{2e}C^{a}_{bc}A_{\mu}^{b}A_{\nu}^{c},
\end{equation}
where $e$, $C^{a}_{bc}$, $A^{a}_{\nu}$s and $\gamma_{ab}=-\frac{\Gamma_{ab}}{|\det \Gamma_{ab}|^{\frac{1}{N}}}$ are the gauge coupling constant,  gauge group structure constant,  gauge potential  and the metric tensor of the gauge group, respectively.  Also, $\Gamma_{ab}=C_{ad}^{c}C_{bc}^{d}$ and $|\det \Gamma_{ab}|>0$. Now,
the black brane solution for this model is given by
\begin{equation}\label{20}
ds^{2}=-\frac{r^{2}N(r)^{2}}{l^{2}}f(r)dt^{2}+\frac{l^{2}dr^{2}}{r^{2}f(r)}+r^{2}h_{ij}dx^{i}dx^{j},
\end{equation}
where metric function is given by
\begin{equation}\label{21}
f(r)=1-\frac{b^{4}}{r^{4}}-\frac{2Q^{2}l^{2}}{r^{4}}\ln r+m^{2}l^{2}\left(\frac{c_{0}c_{1}}{3r}+\frac{c_{0}^{2}c_{2}}{r^{2}}+\frac{2c_{0}^{3}c_{3}}{r^{3}}\right),
\end{equation}
where $N(r)$ is constant by variation of $f(r)$.
By using relation $f(r)|_{r=r_{+}}=0$, the event horizon and value of $b$    can be obtained.  In this case, using Hawking-Bekenstein formula,  Hawking temperature and the entropy are calculated,  respectively, by \cite{13,27},
\begin{equation}
T=  \frac{r_{+}}{4\pi l^{2}}\left(4-\frac{2e^{2}l^{2}}{r_{+}^{4}}-m^{2}l^{2}(\frac{c_{0}c_{1}}{3r_{+}}+\frac{2c_{0}^{2}c_{2}}{r_{+}^{2}}+\frac{6c_{0}^{3}c_{3}}{r_{+}^{3}})\right),\label{22}
\end{equation}
and
\begin{equation}
S=\frac{A}{4G}=\frac{r_{+}^{3}V}{4l^{3}G}. \label{23}
\end{equation}

Now, in order to examine universal relations
for a  perturbatively changing  theory in a manner controlled by a parameter   $\epsilon$, we define action as follows
\begin{equation}
I_\epsilon =\int d^{5}x\sqrt{-g}(R- 2(1+\epsilon)\Lambda-\gamma_{ab}F^{a}_{\mu\nu}F^{b\mu\nu}+m^{2}\Sigma_{i=1}^{4}c_{i}U_{i}(g, f)).\label{24}
\end{equation}
Corresponding to this modified action, we calculate modified thermodynamic quantities such as mass ($M$) and Hawking temperature ($T$)  as follows
\begin{eqnarray}
M &=&\frac{3(1+\epsilon)S^{\frac{4}{3}}}{4^{\frac{4}{3}}\pi^{\frac{4}{3}}lV^{\frac{1}{3}}}+\frac{2m^{2}Q^{2}(2^{\frac{2}{3}}l^{2}\pi^{\frac{1}{3}}S^{\frac{2}{3}}V^{\frac{4}{3}}c_{0}c_{1}+6\times2^{\frac{1}{3}}l\pi^{\frac{2}{3}}S^{\frac{1}{3}}V^{\frac{5}{3}}c_{0}^{2}c_{2}+24\pi V^{2}c_{0}^{3}c_{3})}{l^{4}(1+\epsilon)S},\\\label{25}
T&=&\frac{(1+\epsilon)S^{\frac{1}{3}}}{2^{\frac{2}{3}}l\pi^{\frac{4}{3}}V^{\frac{1}{3}}}-\frac{2m^{2}Q^{2}(2^{\frac{2}{3}}l^{2}\pi^{\frac{1}{3}}S^{\frac{2}{3}}V^{\frac{4}{3}}c_{0}c_{1}+12\times2^{\frac{1}{3}}l\pi^{\frac{2}{3}}S^{\frac{1}{3}}V^{\frac{5}{3}}c_{0}^{2}c_{2}+72\pi V^{2}c_{0}^{3}c_{3})}{3(1+\epsilon)l^{4}S^{2}}. \label{26}
\end{eqnarray}
Electric potential is calculated by
\begin{equation}
\Phi=\frac{4m^{2}Q(2^{\frac{2}{3}}l^{2}\pi^{\frac{1}{3}}S^{\frac{2}{3}}V^{\frac{4}{3}}c_{0}c_{1}+6\times2^{\frac{1}{3}}l\pi^{\frac{2}{3}}S^{\frac{1}{3}}V^{\frac{5}{3}}c_{0}^{2}c_{2}+24\pi V^{0}c_{0}^{3}c_{3}}{(1+\epsilon)l^{4}S}.\label{27}
\end{equation}
We simplify Eq. (\ref{25}) to identify the constant correction parameter $ \epsilon $
as follows
\begin{eqnarray}
 \epsilon &=& -1+\frac{2\times2^{\frac{2}{3}}l^{4}M\pi^{\frac{4}{3}}V^{\frac{1}{3}}}{3l^{3}S^{\frac{4}{3}}}\pm \frac{2^{\frac{5}{3}}}{3l^{3}S^{\frac{7}{3}}}
 \left[ l^{8}M^{2}\pi^{\frac{8}{3}}S^{2}V^{\frac{2}{3}}-6l^{5}m^{2}Q^{2}S^{3}\pi^{\frac{5}{3}}V^{\frac{5}{3}}c_{0}c_{1}\right.\nonumber\\
 &-& \left. 18\times 2^{\frac{2}{3}}l^{4}m^{2}\pi^{2}Q^{2}V^{2}S^{\frac{8}{3}}c_{0}^{2}c_{2}-72\times2^{\frac{1}{3}}l^{3}m^{2}Q^{2}(\pi SV)^{\frac{7}{3}}c_{0}^{3}c_{3}  \right]^{\frac{1}{2}}.\label{28}
\end{eqnarray}
The derivatives of equations (\ref{25}) and (\ref{26}) yield
\begin{equation}
-T\frac{\partial S}{\partial \epsilon}= \frac{\mathcal{X}}{\mathcal{Y}},\label{30}
\end{equation}
where
\begin{eqnarray}
\mathcal{X} &=&- 3l^{\frac{3}{2}} S^{\frac{4}{3}}V^{\frac{1}{3}}\left[3\times2^{\frac{1}{3}}l^{3}S^{\frac{7}{3}}(1+\epsilon)^{2}-4m^{2}\pi^{\frac{5}{3}}Q^{2}V^{\frac{5}{3}}c_{0}\Big\{2^{\frac{2}{3}}l^{2}S^{\frac{2}{3}}c_{1}\right.\nonumber\\
& +&\left.  12l(2\pi SV)^{\frac{1}{3}}c_{0}c_{2}+72(\pi V)^{\frac{2}{3}}c_{0}^{2}c_{3}\Big\}\right]   \left[l^{5}M^{2}\pi+6m^{2}Q^{2}\right.\nonumber\\
&+&\left. S^{\frac{1}{3}}Vc_{0}[-l^{2}S^{\frac{2}{3}}c_{1}-3(2V\pi)^{\frac{1}{3}}c_{0}\{2S^{\frac{1}{3}}lc_{2}+4(\pi V)^{\frac{1}{3}}c_{0}c_{3}\}]\right]^\frac{1}{2},\nonumber\\
\mathcal{Y} &=& 4\times2^{\frac{2}{3}}l^{4}\pi^{\frac{13}{6}}V^{\frac{2}{3}}(1+\epsilon)\bigg[15(lmQS)^{2}V^{\frac{4}{3}}c_{0}c_{1}+54\times2^{\frac{2}{3}}lm^{2}\pi^{\frac{1}{3}}Q^{2}(SV)^{\frac{5}{3}}c_{0}^{2}c_{2}\nonumber\\
&+&252\times2^{\frac{1}{3}}(mQV)^{2}\pi^{\frac{2}{3}}S^{\frac{4}{3}}c_{0}^{3}c_{3}-4\Big(l^{5}M^{2}\pi SV^{\frac{1}{3}}+\sqrt{\pi}lM
\left(l^{3}S^{2}V^{\frac{2}{3}}(l^{5}M^{2}\pi \right.\nonumber\\
&+& \left. 6m^{2}Q^{2}+S^{\frac{1}{3}}Vc_{0}[-l^{2}S^{\frac{2}{3}}c_{1}-3(2V\pi)^{\frac{1}{3}}c_{0}\{2S^{\frac{1}{3}}lc_{2}+4(\pi V)^{\frac{1}{3}}c_{0}c_{3}\}]\right)^\frac{1}{2}  \Big)\bigg].\nonumber
\end{eqnarray}
 Now, in order to check universal relation, we need mass at extremal point which
 corresponds to vanishing temperature.
 Here, the expression for extremal mass is calculated by
\begin{equation}
M_{ext}=\frac{\mathcal{A}+\mathcal{B}  }{\mathcal{C}}, \label{31}
\end{equation}
where
\begin{eqnarray}
\mathcal{A}&=&3l^{\frac{12}{5}}V^{\frac{2}{3}}(1+\epsilon)\Big(\frac{(mQ)^{\frac{6}{5}}V(c_{0}c_{1})^{\frac{3}{5}}}{l^{\frac{3}{5}}(1+\epsilon)^{\frac{6}{5}}}\Big)^{\frac{4}{3}},\nonumber\\
\mathcal{B}&=&c_{1}^{-\frac{3}{5}}\Big(12(mQ)^{\frac{4}{5}}V^{\frac{4}{3}}(1+\epsilon)^{\frac{1}{5}}c_{0}^{\frac{2}{5}}\Big[l^{2}\Big(\frac{(mQ)^{\frac{6}{5}}V(c_{0}c_{1})^{\frac{3}{5}}}{l^{\frac{3}{5}}(1+\epsilon)^{\frac{6}{5}}}\Big)^{\frac{2}{3}}+c_{1}\nonumber\\
&+&3\times6^{\frac{1}{5}}V^{\frac{1}{3}}c_{0}\Big\{l\Big(\frac{(mQ)^{\frac{6}{5}}V(c_{0}c_{1})^{\frac{3}{5}}}{l^{\frac{3}{5}}(1+\epsilon)^{\frac{6}{5}}}\Big)^{\frac{1}{3}}c_{2}+2\times6^{\frac{1}{5}}V^{\frac{1}{3}}c_{0}c_{3}\Big\}\Big]\Big), \nonumber\\
\mathcal{C}&=&6^{\frac{4}{5}}l^{\frac{17}{5}}V.\nonumber
\end{eqnarray}
From the expressions  (\ref{30}) and (\ref{31}), we observe that
\begin{equation}\label{32}
\frac{\partial M_{ext}}{\partial\epsilon}\neq -T\frac{\partial S}{\partial \epsilon}.
\end{equation}
The above expression   confirms that the equality of universal relation does not hold. So, the universal relationship is violated for this model as well. Here one can also check for another universal relation that is valid of black holes.
To do so, we compute
\begin{equation}
\frac{\partial \epsilon}{\partial Q}=-\frac{4m^{2}Qc_{0}\pi^{\frac{5}{6}}V^{\frac{4}{3}}\left[(2S)^{\frac{2}{3}}l^{2}c_{1}+6l(2\pi VS)^{\frac{1}{3}}c_{0}c_{2}+24(\pi V)^{\frac{2}{3}}c_{0}^{2}c_{3}\right]}{S\sqrt{l^{3}  (l^{5}M^{2}\pi+6m^{2}Q^{2}S^{\frac{1}{3}}Vc_{0}(-l^{2}S^{\frac{2}{3}}c_{1}-3(2\pi V)^{\frac{1}{3}}c_{0}((2S)^{\frac{1}{3}}lc_{2}+4(\pi V)^{\frac{1}{3}}c_{0}c_{3})))}}.\label{33}
\end{equation}
Exploiting expressions (\ref{27}), (\ref{30}), (\ref{31}) and (\ref{33}), we have
\begin{equation}\label{34}
\begin{split}
&-\Phi\frac{\partial Q}{\partial \epsilon}\neq \frac{\partial M_{ext}}{\partial\epsilon}\neq -T\frac{\partial S}{\partial \epsilon}.
\end{split}
\end{equation}
Remarkably, we observe that this universal relation is also violated for the Einstein-Yang-Mills AdS black brane in massive gravity.
As it has been shown so far, universal relations compatible with black holes are completely violated in the black brane structure. The interesting thing is that if we assume the coefficients $ c_{i} = 0 $, each of the above equations (\ref{34})  tends to infinity. As a result, with a series of straightforward calculations and simplifications, we get the same values for each expression.
Eventually, under such assumption, we can also create this universal relationship for the Einstein-Yang-Mills AdS black brane in massive gravity as follows
\begin{equation*}
\begin{split}
&-\Phi\frac{\partial Q}{\partial \epsilon}=\frac{\partial M_{ext}}{\partial\epsilon}= -T\frac{\partial S}{\partial \epsilon}\approx \frac{m^{\frac{8}{5}}Q^{\frac{8}{5}}V^{\frac{2}{3}}}{l^{\frac{9}{5}}(1+\epsilon)^{\frac{12}{5}}}.
\end{split}
\end{equation*}
The crucial point about this model of black brane is that due to the charge, the conjecture derived from string theory, viz., the weak gravity conjecture, can be examined. In the literature, weak gravity conjecture is defined as the gravity is the weakest force.  Weak gravity conjecture states that the charge-to-mass ratio is greater than one (i.e. $\frac{Q}{M}> 1$). The swampland program and weak gravity conjecture have always been studied in many past works, such as the study of inflation models, the structure of black holes, and other cosmological implications, etc., in the Refs. \cite{a,e, 28, 30, 32,33,34,35,36,37}.
Now we want to evaluate another of these implications about the black brane from the plot.
\begin{figure}[h!]
\begin{center}
\subfigure[]{
\includegraphics[height=6cm,width=6cm]{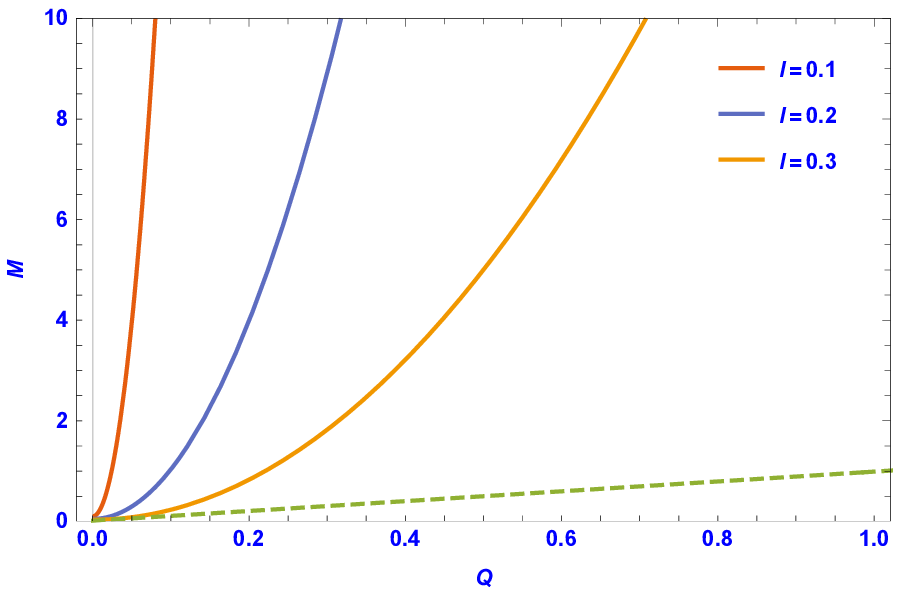}
\label{1a}}
\subfigure[]{
\includegraphics[height=6cm,width=6cm]{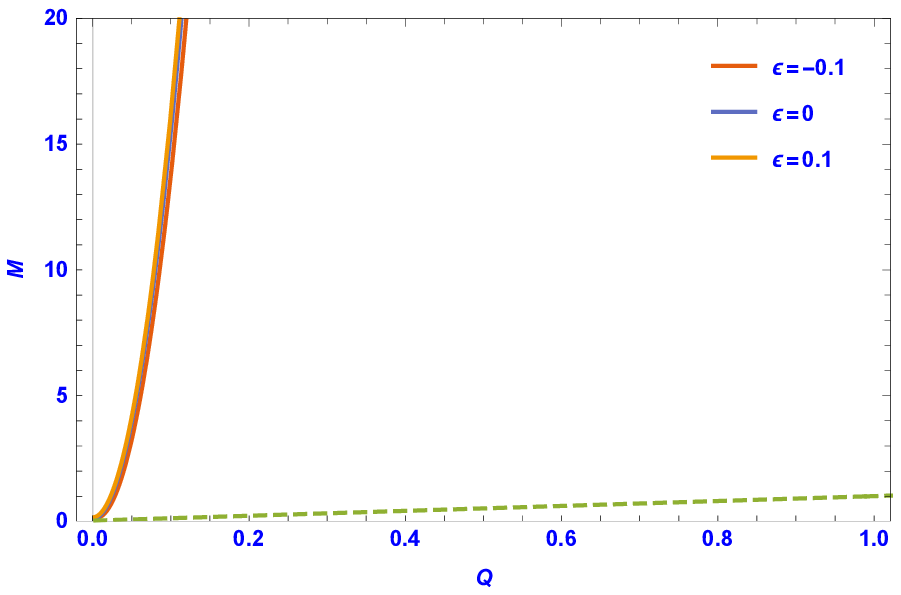}
\label{1b}}
\caption{\small{(a) The plot of unmodified $M$ in terms of $Q$ with   $l=0.1, 0.2, 0.3$  and (b) the plot of modified $M$ in term of $Q$ with  $l=0.1$ and $\epsilon=-0.1, 0, 0.1$ in (b). }}
\label{01}
\end{center}
\end{figure}
With all the above mentioned explanations, we now want to compare unmodified and modified mass as depicted in Fig. \ref{01}. Here, we assumed some parameters to have constant value and plot the mass in terms of charge $Q$. The dashed lines show states     when the mass-to-charge ratio is one. It can be seen from plot that the mass ratio of the unmodified black brane is more than one. The AdS space radius is assumed to have values  $l= 0.1, 0.2, 0.3$. The changes can be seen in figure \ref{1a}. However, in figure \ref{1b}, we consider the modified state for the mass concerning $l=0.1$ and different values $\epsilon$. As shown in figure \ref{1b}, for positive constant correction   mass increases; however for negative constant correction,  the mass decreases. Of course, in the numerical calculation of entropy, there exists a contrast. This means that entropy decreases with positive correction. In fact, according to the above mentioned  concepts,  for negative small correction, the mass of the black hole decreases to one, and the charge to mass ratio increases or the mass charge ratio decreases, which are quite satisfied by the weak gravity conjecture.

\section{Perturbative General Anisotropic Black Brane in Horava-Lifshitz  Gravity}\label{sec4}

In this section, we study universal relations for an anisotropic brane model. Various theories are used to describe such system. One of them is Horava-Lifshitz and Einstein-Hilbert gravity with a scalar field \cite{38,39}. There are several ways to develop Horava-Lifshitz gravity theory  \cite{40,41}.  Here, we consider  a
five-dimensional axion-dilaton-gravity action
\begin{equation}
I=\int d^{5}x\sqrt{-g}\left[R-2\Lambda-\frac{1}{2}(\partial \phi)^{2}-\frac{1}{2}\exp(2\phi)(\partial \chi)^{2}\right],\label{35}
\end{equation}
where $\chi$ is axion field and $\phi$ is dilaton field. The   black brane solution for this theory  is described by following metric:
\begin{equation}
ds^{2}=-r^{2\alpha}h(r)dt^{2}+\frac{dr^{2}}{r^{2}f(r)}+\frac{r^{2}}{l^{2}}b(r)(dx^{2}+dy^{2})+\frac{r^{2}}{l^{2}}k(r)dz^{2},\label{36}
\end{equation}
where $h(r)$ and $f(r)$ are the blackening factors and $\alpha$ is the dynamical critical exponent. The above metric (\ref{36}) corresponding to dimensionless coordinate $(u=\frac{r_{+}^{2}}{r^{2}})$ takes the following form:
\begin{equation}
ds^{2}=-\frac{r_{+}^{2\alpha}}{u^{\alpha}l^{2\alpha}}H(u)dt^{2}+\frac{l{2}du^{2}}{4u^{2}F(u)}+\frac{r_{+}^{2}}{ul^{2}}B(u)(dx^{2}+dy^{2})+\frac{r_{+}^{2}}{ul^{2}}K(u)dz^{2}.\label{37}
\end{equation}
From the above expression, it is obvious that the solution is isotropic in the $xy$-directions but not in $z$-direction \cite{14}.

Corresponding to the above solution, the Hawking temperature and the Hawking-Bekenstein entropy density are given, respectively, by  \cite{42}
\begin{equation}
T=\frac{r_{+}^{\alpha}}{2\pi l^{\alpha+1}}\sqrt{F^{'}H^{'}}|_{u=1}=\frac{r_{+}^{\alpha}}{2\pi l^{\alpha+1}}\sqrt{\frac{F}{H}}H^{'}|_{u=1},\label{38}
\end{equation}
and
\begin{equation}
s=\frac{4\pi}{V}\int d^{3}x\sqrt{-g}=4\pi(\frac{r_{+}}{l})^{3}B(u=1)\sqrt{K(u=1)}.\label{39}
\end{equation}
In order to study the universal relation for this model, we first modify the  action by introducing a small correction parameter as follows
\begin{equation}
I_\epsilon =\int d^{5}x\sqrt{-g}\left[R-2(1+\epsilon) \Lambda-\frac{1}{2}(\partial \phi)^{2}-\frac{1}{2}\exp(2\phi)(\partial \chi)^{2}\right].\label{40}
\end{equation}
It is matter of calculation to obtain  modified thermodynamic mass and temperature
corresponding to action (\ref{40}). These are, respectively,
\begin{equation}
M=\frac{3\times2^{-1-\frac{2\alpha}{3}}\sqrt{FH}l^{-1-\alpha}\pi^{-1-\frac{\alpha}{3}}S\left(\frac{lS^{\frac{1}{3}}(1+\epsilon)}{(BV)^{\frac{1}{3}}K^{\frac{1}{6}}}\right)^{\alpha}}{3+\alpha},\label{41}
\end{equation}
and
\begin{equation}\label{42}
T=2^{-1-\frac{2\alpha}{3}}\sqrt{FH}l^{-1-\alpha}\pi^{-1-\frac{\alpha}{3}}
\left(\frac{lS^{\frac{1}{3}}(1+\epsilon)}{(BV)^{\frac{1}{3}}K^{\frac{1}{6}}}\right)^{\alpha}.
\end{equation}
Also, upon solving equation (\ref{41}), we get the constant correction parameter $\epsilon$  as follows
\begin{equation}\label{43}
\epsilon=-1+\frac{3^{\frac{-1}{\alpha}}B^{\frac{1}{3}}K^{\frac{1}{6}}V^{\frac{1}{3}}(\frac{2^{1+{\frac{2\alpha}{3}}}(3+\alpha)l^{1+\alpha}\pi^{1+{\frac{\alpha}{3}}}}{\sqrt{FH}S})^{\frac{1}{\alpha}}}{lS^{\frac{1}{3}}}.
\end{equation}
The derivative of $\epsilon$ with respect to entropy is computed as
\begin{equation}\label{44}
\frac{\partial\epsilon}{\partial S}=-\frac{(3+\alpha)(BV)^{\frac{1}{3}}K^{\frac{1}{6}} \left(\frac{lS^{\frac{1}{3}}(1+\epsilon)}{(BV)^{\frac{1}{3}}K^{\frac{1}{6}}}\right)
}{3\alpha l S^{\frac{4}{3}}}.
\end{equation}
This eventually leads to
\begin{equation}
 -T\frac{\partial S}{\partial \epsilon}
= \frac{3\times2^{-1-{\frac{2\alpha}{3}}}(BV\sqrt{K})^{\frac{4}{3}}\alpha\frac{\sqrt{FH}}{l^{4}(1+\epsilon)^{4}}l^{-\alpha}\pi^{-1-{\frac{\alpha}{3}}} \left(\frac{l(BV\sqrt{K})^{\frac{1}{3}}(1+\epsilon)}{l(1+\epsilon)(BV)^{\frac{1}{3}}K^{\frac{1}{6}}}\right)^{\alpha-1}  }{(3+\alpha)(BV)^{\frac{1}{3}}K^{\frac{1}{6}}}.\label{45}
\end{equation}
Following the methodology of previous sections, in order to check the   universal relation, we   first evaluate $S$ by taking  $T=0$  and then plug it's value in equation (\ref{41}) to get extremal mass
\begin{equation}\label{46}
M_{ext}=\frac{3\times2^{-1-{\frac{2\alpha}{3}}}B\sqrt{FHK}l^{-4-\alpha}\pi^{-1-{\frac{\alpha}{3}}}V\left(\frac{l(1+\epsilon)(\frac{BV\sqrt{K}}{l^{3}(1+\epsilon)^{3}})^{\frac{1}{3}}}{B^{\frac{1}{3}}K^{\frac{1}{6}}}\right)^{\alpha}}{(3+\alpha  )(1+\epsilon)^{3}},
\end{equation}
 The derivative of $M_{ext}$ with respect to $\epsilon$ gives
 \begin{equation}\label{47}
\frac{\partial M_{ext}}{\partial\epsilon}=-\frac{9\times2^{-1-{\frac{2\alpha}{3}}}B\sqrt{FHK}l^{-4-\alpha}\pi^{-1-{\frac{\alpha}{3}}}V\left(\frac{l(1+\epsilon)(\frac{BV\sqrt{K}}{l^{3}(1+\epsilon)^{3}})^{\frac{1}{3}}}{B^{\frac{1}{3}}K^{\frac{1}{6}}}\right)^{\alpha}}{(3+\alpha)(1+\epsilon)^{4}}.
\end{equation}
Clearly, it is evident that   equations (\ref{45}) and (\ref{47})  are not equal,
i.e., $$-T\frac{\partial S}{\partial \epsilon}\neq \frac{\partial M_{ext}}{\partial\epsilon}.$$
However, remarkably,  if we consider $-1<\alpha<0$ in equations (\ref{45}) and (\ref{47}), a universal relation will be established as these two equations will become almost the same. Thus, we have
 \begin{equation}\label{48}
 \begin{split}
-T\frac{\partial S}{\partial \epsilon}=\frac{\partial M_{ext}}{\partial\epsilon}\approx-\delta\frac{\sqrt{FH}\sqrt{K}V}{l^{3.}(1+\epsilon)^{4}},
\end{split}
\end{equation}
where $\delta$ is a constant coefficient.

Here  a new universal relation is obtained for the black brane solution. Many researchers have done different works according to different conditions and proved this universal relation in different ways.
At the end,  we conclude that the universal relations, which are valid for black holes, are not valid for black brane or violates in any way. This is an interesting result. Of course present study end-up with a question that is there any way or  conditions  to establish these universal relations or not?

\section{Conclusions}\label{sec5}

Researchers recently studied universal thermodynamic (entropy/extremality) relations   by modifying
general relativity for various black holes such  as rotating-AdS black holes, charged black holes, rotating black holes in massive gravity, etc. Here, the main concern of the paper is to extend such  study  of universal thermodynamic relations for the case of black brane solutions.
Therefore, we have
considered several different structures of the black brane such as black brane solution in
Rastall AdS massive gravity, Einstein-Yang-Mills AdS black brane solution in massive
gravity, and general anisotropic black brane in Horava-Lifshitz gravity.
Furthermore, we  obtained the modified thermodynamic relations by considering a small constant correction added to the black brane action. Then, by analytical calculations, we have computed   universal thermodynamic extremality relation by
the perturbative corrections of generic thermodynamic systems. In this connection, we
have observed   remarkably that, in general, these universal relations are not valid  for different black branes with various properties and structures. However, under certain
approximation, such relations may be valid for black branes also. This is an important result.\\
 For future perspective, it will be interesting to study black holes with other structures such as Einstein-Gauss-Bonnet along with features such as higher dimensions taking into account which may provide exciting results.

\end{document}